\documentclass[prl,reprint,twocolumn,showpacs,preprintnumbers,amsmath,amssymb,superscriptaddress,floatfix]{revtex4-2}
\usepackage[T1]{fontenc}
\usepackage[ansinew]{inputenc}
\usepackage{graphics}
\usepackage{dcolumn}
\usepackage{bm}
\usepackage{amsthm}
\usepackage{amsmath}
\usepackage{amssymb}
\usepackage{diagbox}
\usepackage{hyperref}
\usepackage{url}
\usepackage{xcolor}

\def\afflux{Department of Physics and Materials Science, University of Luxembourg, 162A~Avenue de la Faiencerie, L-1511 Luxembourg, Grand Duchy of Luxembourg}

\begin{document}

\title{Framework for Polarized Magnetic Neutron Scattering from Nanoparticle Assemblies with Vortex-Type Spin Textures}

\author{Michael P.\ Adams}\email[Electronic address: ]{michael.adams@uni.lu}
\affiliation{\afflux}

\author{Evelyn P.\ Sinaga}
\affiliation{\afflux}

\author{\v{S}tefan Li\v{s}\v{c}\'{a}k}
\affiliation{\afflux}

\author{Andreas Michels}\email[Electronic address: ]{andreas.michels@uni.lu}
\affiliation{\afflux}

%%%%%%%%%%%%%%%%%%%%%%%%%%%%%%%%%%%%%%%%%%%%%%%%%%%%%%%%%%%%%%%%%%%

\begin{abstract}
Within the framework of the recently introduced multi-nanoparticle power-series expansion method for the polarized small-angle neutron scattering (SANS) cross section, we present analytical expressions for the polarized SANS observables arising from dilute nanoparticle assemblies with antisymmetric vortex-type spin structures. We establish connections between the magnetic correlation coefficients and the magnetic field-dependent vortex-axes distribution function, which is related to the random orientations of the magnetocrystalline anisotropy axes of the nanoparticles. Our analytical results are validated through a comparative analysis with micromagnetic simulations. This framework contributes to a comprehensive understanding of polarized magnetic neutron scattering from spherical nanoparticle systems exhibiting vortex-type spin structures.
\end{abstract}

\date{\today}

\maketitle

%%%%%%%%%%%%%%%%%%%%%%%%%%%%%%%%%%%%%%%%%%%%%%%%%%%%%%%%%%

\textit{Introduction.} Magnetic nanoparticles, which are in the scope of immense interdisciplinary research, offer versatile applications, e.g., in materials science, nanotechnology, and biomedicine~\cite{de2008applications,diebold2010applications,baetke2015applications,stark2015industrial,han2019applications,lakbenderdisch2021,BATLLE2022}. They open up new possibilities in the nanoscopic realm and drive technological advances and breakthrough discoveries. But still, at the current stage of research, it is an immense challenge to characterize their internal spin structure, which is generally to be expected as nonuniform (e.g., \cite{peddis2017,lappas2019,zakutna2020,leighton2020,kryckaprm2020,koehler2021nanoscale,iglesias2021,eli2023,zakutna2023}).

Magnetic small-angle neutron scattering (SANS) is possibly the only technique to probe the spatial variation of spin structures on a scale of $\sim$$1$$-$$100 \, \mathrm{nm}$ and in the bulk of the material~\cite{rmp2019,michelsbook}. Recent advances in the understanding of magnetic SANS from complex nanoparticle systems have been achieved by the marriage of micromagnetic theory and magnetic neutron scattering formalism, through both computer simulations and analytical calculations~\cite{metmi2016,michelsPRB2016,mirebeau2018,mistonov2019,laura2020,metlov2022,arena2023,ukleev2024,evelynprb2024}. Although computer simulations offer considerable potency in predicting neutron scattering observables for intricate nanoparticle assemblies, their drawback lies in their time-intensive nature, vast parameter space, and the inherent challenge of interpreting results. This complexity hinders the derivation of overarching conclusions and poses a substantial obstacle in formulating generalized statements.

To address these challenges, Adams~et~al.~\cite{adamsprb2024} introduced the multi-nanoparticle power-series expansion (MNPSE) method to study the neutron scattering signatures from spherical nanoparticle assemblies featuring diverse types of magnetic surface anisotropy. Here, we use the MNPSE approach to predict the main features of nanoparticle assemblies with inherent {\it vortex-type spin textures} as to be seen in the neutron scattering observables. Vortex-type structures are ubiquitous in magnetism research and are encountered in many systems, such as in type-II superconductors~\cite{forgan2011}, GdCo$_2$ micropillars~\cite{metlovnp2021}, Nd-Fe-B magnets~\cite{bersweilerprb2023}, iron oxide nanoparticles~\cite{samardak2016,usov2018,knobel2023} and nanoflowers~\cite{BATLLE2024}, or the very recently discovered topological vortex rings in a chiral magnetic nanocylinder~\cite{nagaosaprl2024}. Our results, which replace the conventional analytical formulation for the superspin model, enable the straightforward prediction of the spin-flip SANS cross section $I_{\mathrm{sf}}(q)$ and the corresponding spin-flip pair-distance distribution function $p_{\mathrm{sf}}(r)$ arising from spatially antisymmetric spin structures, such as nanovortices, through easily-applicable analytical expressions.

The Letter is organized as follows: We start out by analyzing the main features of the first-order MNPSE method for the spin-flip SANS cross section and the pair-distance distribution function. This approach is valid for an {\it arbitrary linear magnetization distribution}. Subsequently, for the particular case of a {\it linear vortex}, we derive analytical expressions for the two- and one-dimensional SANS observables. The analytical expressions are compared to the results of micromagnetic computations. We refer to the Supplemental Material~\cite{michael2024sm} for details regarding the analytical derivations and the micromagnetic SANS simulations.

\begin{figure*}[tb!]
\centering
\resizebox{2.0\columnwidth}{!}{\includegraphics{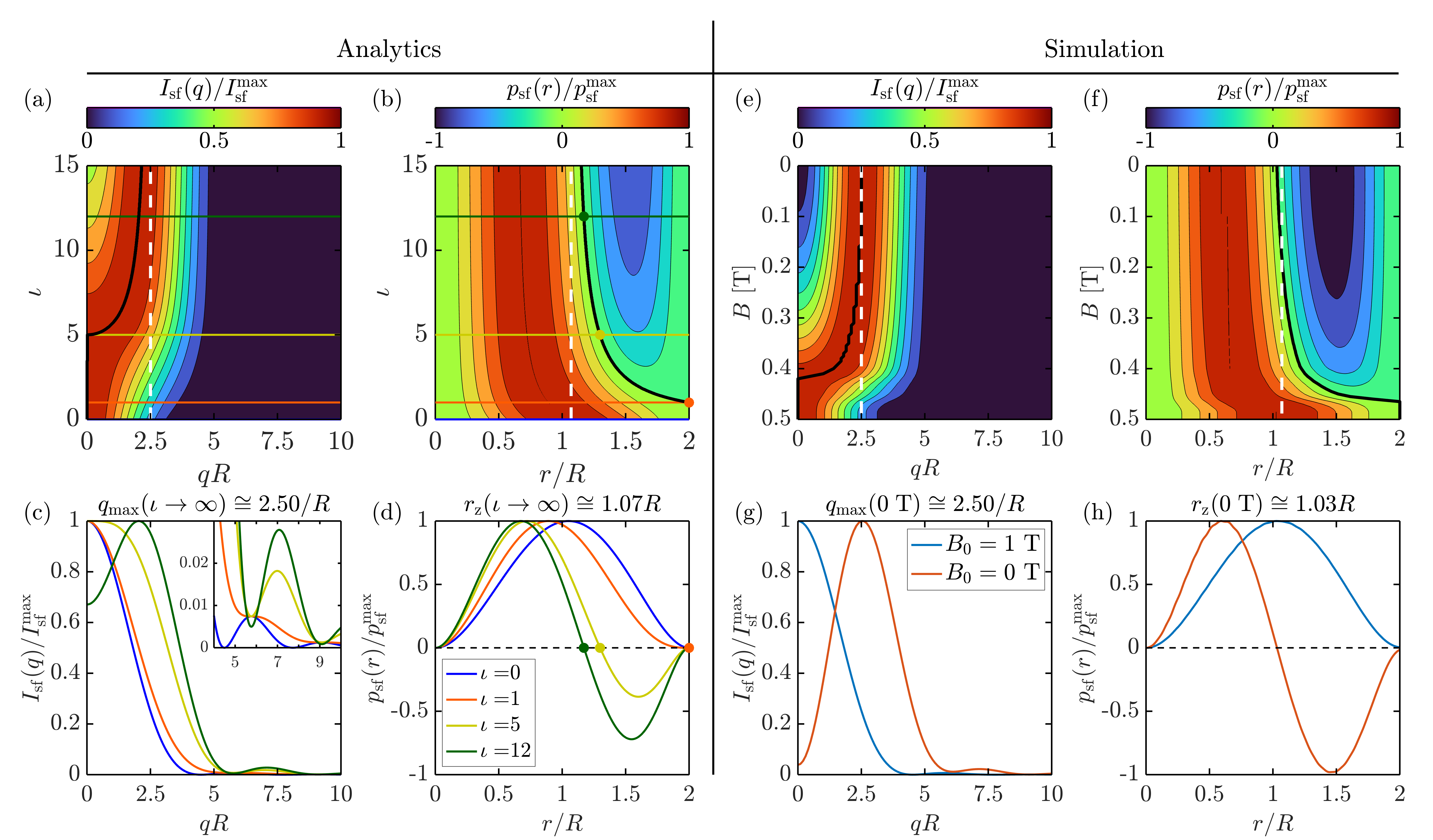}}
\caption{``Phase diagram'' for the azimuthally-averaged spin flip SANS cross section $I_{\mathrm{sf}}(q)$ [Eq.~\eqref{eq:AzimuthallyAveragedSANScrossSectionFirstOrderModel}] and for the spin-flip pair-distance distribution function $p_{\mathrm{sf}}(r)$ [Eq.~\eqref{eq:FirstOrderPairDistanceDistributionFunction}] within the limits of the first-order magnetization model. The left panel shows the analytical results, while the right panel features the corresponding results of the micromagnetic simulations. The ratio $\iota$ of the zero-order coefficient $I_{\mathrm{sf}}^{0}$ and the first-order coefficient $I_{\mathrm{sf}}^{1}$ determines the appearance of vortex-type spin structures. Field ($B_0 = \mu_0 H_0$) variations in the simulations correspond to $\iota$~variations in the analytical part (zero field: $\iota \rightarrow \infty$; saturation: $\iota \rightarrow 0$). (a)~Color-coded plot of the normalized $I_{\mathrm{sf}}(q)$ as a function of $\iota = I_{\mathrm{sf}}^{1} / I_{\mathrm{sf}}^{0}$ and $qR$. The black solid line in (a) describes the shift of the maximum in $I_{\mathrm{sf}}(q)$ towards $q_{\mathrm{max}} \cong 2.50/R$ [white dashed line, compare (e)]. (b)~Normalized $p_{\mathrm{sf}}(r)$ as a function of $\iota$ and $r/R$. The black solid line in (b) describes the shift of the zero in $p_{\mathrm{sf}}(r)$ towards $r_{\mathrm{z}} \cong 1.07 R$ [white dashed line, compare (f)]. (c)~Normalized $I_{\mathrm{sf}}(q R)$ and (d) normalized $p_{\mathrm{sf}}(r/ R)$ for different $\iota$ [see inset in (d)]; the inset in (c) displays $I_{\mathrm{sf}}(q)/I_{\mathrm{sf}}^{\mathrm{max}}$ for $4 < qR < 10$. The colored horizontal lines in (a) and (b) correspond, respectively, to the curves in (c) and (d).}
\label{fig1}
\end{figure*}

\textit{Linear MNPSE method.} This approach is based on the following expansion for the magnetization vector field,
\begin{align}
    \mathbf{M}'(\mathbf{r}') = 
    \begin{bmatrix}
        m_0^{x}   
        \\
        m_0^{y}   
        \\
        m_0^{z}   
    \end{bmatrix}
    +
    \begin{bmatrix}
        m_{1}^{xx} & m_{1}^{xy} & m_{1}^{xz}
        \\
        m_{1}^{yx} & m_{1}^{yy} & m_{1}^{yz}
        \\
        m_{1}^{zx} & m_{1}^{zy} & m_{1}^{zz}
    \end{bmatrix}
    \cdot
    \begin{bmatrix}
        x' 
        \\
        y' 
        \\
        z'
    \end{bmatrix} , \label{eq:LinearMagnetizationFunction}
\end{align}
where $\mathbf{r}' = [x', y', z']$ denotes the position vector in the local particle frame. The model consists of 12 expansion coefficients per particle, i.e., 3 zero-order coefficients $m_0^{i}$ and 9 first-order coefficients $m_1^{jk}$. For a dilute assembly of spherical nanoparticles, the MNPSE formalism yields the following expression for the azimuthally-averaged spin-flip SANS cross section~\cite{michael2024sm,adamsprb2024}:
\begin{align}
    I_{\mathrm{sf}}(q) &= I_{\mathrm{sf}}^{0} [f(qR)]^2 +
    I_{\mathrm{sf}}^{1} [ f'(qR) ]^2,
\label{eq:AzimuthallyAveragedSANScrossSectionFirstOrderModel}
\end{align}
where $R$ is the particle radius, and the field-dependent coefficients $I_{\mathrm{sf}}^{0}$ and $I_{\mathrm{sf}}^{1}$ represent complicated averages of the magnetization coefficients $m_0^i$ and $m_1^{jk}$ over the particle assembly and over the detector plane. The corresponding basis functions are given by ($u=qR$):
\begin{align}
    f(u) &= \frac{\sin u - u \cos u}{u^3} \nonumber ,
    \\
    f'(u) &= \frac{d f}{du} = \frac{(u^2 - 3) \sin u + 3 u \cos u}{u^4} \nonumber .
\end{align}
Here, $f(u)$ is the form-factor function of the unit sphere~\cite{michelsbook} and $f'(u)$ is the related first-order derivative. By inverse Fourier transformation we find from Eq.~\eqref{eq:AzimuthallyAveragedSANScrossSectionFirstOrderModel} the related pair-distance distribution function~\cite{michael2024sm}: 
\begin{align}
\label{eq:FirstOrderPairDistanceDistributionFunction}
p_{\mathrm{sf}}(r) &= I_{\mathrm{sf}}^{0} \frac{\pi r^2}{6 R^3} \left[ 1 - \frac{3r}{4R} + \frac{r^3}{16 R^3} \right] \\
&+  I_{\mathrm{sf}}^{1} \frac{\pi r^2}{10 R^3} \left[ 1 - \frac{5r}{4R} + \frac{5r^3}{16R^3} - \frac{r^5}{32R^5} \right] \nonumber .
\end{align}
While the zero-order contribution ($I_{\mathrm{sf}}^{0}$) arises from symmetric (parallel, positive) correlations only, the first-order contribution ($I_{\mathrm{sf}}^{1}$) contains antisymmetric (antiparallel, negative) correlations. By comparison to micromagnetic simulations using Mumax3~\cite{mumax3new,Leliaert_2018}---including isotropic exchange, a random cubic anisotropy~\footnote{In the numerical micromagnetic simulations we use the materials parameters for (cubic) iron, and we also consider the case of a random uniaxial anisotropy (see \cite{michael2024sm}).}, the Zeeman interaction, and the demagnetizing field---we find that this linear approach [Eqs.~\eqref{eq:AzimuthallyAveragedSANScrossSectionFirstOrderModel} and \eqref{eq:FirstOrderPairDistanceDistributionFunction}] already captures the main features of vortex-type spin textures seen in the SANS observables.

The results that are embodied by Eqs.~\eqref{eq:AzimuthallyAveragedSANScrossSectionFirstOrderModel} and \eqref{eq:FirstOrderPairDistanceDistributionFunction} are summarized in Fig.~\ref{fig1}. Prominent features regarding vortex-type spin structures are the decreased spin-flip scattering intensity $I_{\mathrm{sf}}(q)$ at momentum transfer $q=0$ [see Fig.~\ref{fig1}(a,c,e,g)] and the damped oscillatory behavior of the pair-distance distribution function $p_{\mathrm{sf}}(r)$ exhibiting negative (antiparallel) correlations related to a vortex (see Fig.~\ref{fig1}(b,d,f,h), compare to \cite{laura2020,evelynprb2023,bersweilerprb2023}). In the limiting case of $\iota = I_{\mathrm{sf}}^{1}/I_{\mathrm{sf}}^{0} \rightarrow \infty$ (modeling the remanent state) our linear theory predicts a maximum of $I_{\mathrm{sf}}(q)$ at $q_{\mathrm{max}} \cong 2.50/R$ [maximum of $(f'(u))^2$]. This prediction is in excellent agreement with the result from our micromagnetic simulations, where we find $q_{\mathrm{max}}(B_0 = 0\; \mathrm{T}) \cong 2.50/R$ [see Fig.~\ref{fig1}(g)]. Furthermore, the relevant zero of $p_{\mathrm{sf}}(r)$ is predicted as the result of the following cubic equation that is derived from Eq.~\eqref{eq:FirstOrderPairDistanceDistributionFunction}:
\begin{align}
\nu^3 + 4 \nu^2 + \left( 2 - \frac{10}{3\iota} \right) \nu - \left( 8 + \frac{40}{3\iota} \right) = 0 ,
\label{eq:CubicZeroValueEquation}
\end{align}
where $\nu = r/R$. For $\iota \rightarrow \infty$, Eq.~\eqref{eq:CubicZeroValueEquation} predicts the zero at $r_{\mathrm{z}} \cong 1.07 R$, whereas in our micromagnetic simulations we find $r_{\mathrm{z}}(B_0 = 0\; \mathrm{T}) \cong 1.03 R$ [see Fig.~\ref{fig1}(h)].

Beyond these limits for the momentum transfer $q_{\mathrm{max}}$ and the ``zero'' correlation length $r_{\mathrm{z}}$ we find two specific transition points for $\iota$ in the two-dimensional (2D) maps shown in Fig.~\ref{fig1}(a) and (b). In Fig.~\ref{fig1}(b) we observe for $\iota > 1$ the occurrence of negative correlations [negative values of $p_{\mathrm{sf}}(r)$], and in Fig.~\ref{fig1}(a) we see that the scattering intensity at the origin of reciprocal space, $I_{\mathrm{sf}}(q=0)/I_{\mathrm{sf}}^{\mathrm{max}}$, is constant for $\iota < 5$ and decreases for $\iota > 5$. The micromagnetic simulation results reveal an analogous behavior [compare Fig.~\ref{fig1}(e) and (f)].
 
\begin{figure*}[tb!]
\centering
\resizebox{1.70\columnwidth}{!}{\includegraphics{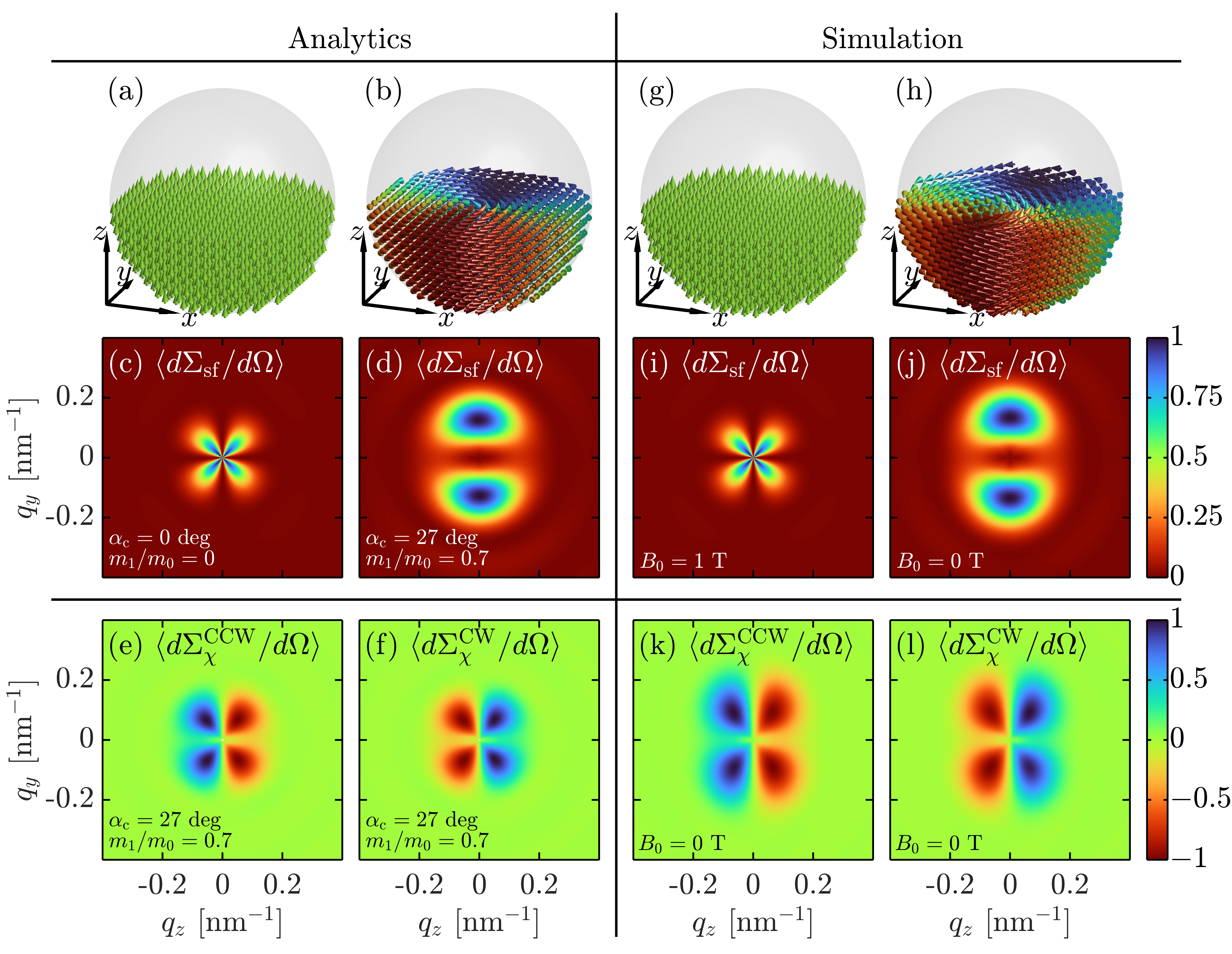}}
\caption{Illustration of the 2D (normalized) spin-flip SANS cross section and chiral function computed from Eqs.~\eqref{eq:SpinFlip2DSANScrossSection} and \eqref{eq:2DChiralSANScross_sectionCCW} reflecting the saturation and remanence cases (particle size: $D = 2R = 40 \, \mathrm{nm}$) (linear color scale). (a,b) and (g,h) show snapshots of the underlying real-space spin structures. The left panel shows the analytical results, while the right panel features the results of the micromagnetic simulations. The incoming neutron beam ($\parallel \mathbf{e}_x$) is perpendicular to the applied magnetic field $\mathbf{H}_0 \parallel \mathbf{e}_z$ ($B_0 = \mu_0 H_0$). The maximum of the spots in (d) and (j) are found at $q_{y,\mathrm{max}} \cong 2.50/R$. (e,f) and (k,l) display the respective chiral functions in the remanent state for counterclockwise (CCW) and clockwise (CW) vortex rotations. Note that the specific values for $\alpha_{\mathrm{c}} = 27^{\circ}$ and for the ratio $m_1/m_0 = 0.7$ in (d) are based on a fit of the analytical function [Eq.~\eqref{eq:SpinFlip2DSANScrossSection}] to the 2D simulation data shown in (j).}
\label{fig2}
\end{figure*} 

\textit{MNPSE method: The case of a linear vortex.} In the formulation of the linear MNPSE method the parameters $I_{\mathrm{sf}}^{0}$ and $I_{\mathrm{sf}}^{1}$ [in Eqs.~\eqref{eq:AzimuthallyAveragedSANScrossSectionFirstOrderModel} and \eqref{eq:FirstOrderPairDistanceDistributionFunction}] are arbitrary functions of the 12 magnetization expansion coefficients in Eq.~(\ref{eq:LinearMagnetizationFunction}). In the following, we aim to adapt the linear MNPSE method to include physically motivated parameters (replacing $I_{\mathrm{sf}}^{0}$ and $I_{\mathrm{sf}}^{1}$). This approach allows us to obtain a scattering model that is closer related to the underlying micromagnetic Hamiltonian in the sense that it contains information on the vortex helicity, on the orientation distribution of the vortex axes, and on the transformation behavior of the energies in the Hamiltonian under space inversion.

We consider a dilute assembly of noninteracting spherical nanoparticles that are rigidly embedded in a homogeneous and nonmagnetic matrix. Each particle is assumed to have a random orientation of its (cubic or uniaxial) magnetic anisotropy axis with respect to the externally applied magnetic field $\mathbf{H}_0 \parallel \mathbf{e}_z$, which defines the laboratory frame of reference. In addition to magnetic anisotropy and the Zeeman interaction, we consider an isotropic exchange energy and, most importantly, the magnetodipolar interaction (see the Supplemental Material~\cite{michael2024sm}). When the spin structure of such a spherical nanoparticle is computed starting from saturation, we always find---using the materials parameters of iron---a vortex-type texture at low fields and for particle sizes larger than about $20 \, \mathrm{nm}$~\cite{laura2020,evelynprb2023}. It is the dipolar interaction that is responsible for the vortex formation.

Based on these simulation results, and with the aim to obtain an approximate expression for the spin-flip SANS cross section of an ensemble of vortex-carrying randomly-oriented nanoparticles, we introduce a magnetization model with a uniform (constant) part of magnitude $m_0$ and a linear vortex term of magnitude $m_1$. More specifically, the basic magnetization vector field is written as:
\begin{align}
\mathbf{M}'(\mathbf{r}') = m_0 \mathbf{e}_z' + m_1 \mathbf{v}(\mathbf{r}') ,
\label{eq:LinearVortexModel}
\end{align}
where $\mathbf{e}_z'= [0, 0, 1]$ is the unit vector in $z'$~direction, $\mathbf{v}(\mathbf{r}') = [-y', x', 0]$ is the linear vortex field, and $\mathbf{r}' = [x',y',z']$ is the position vector with reference to the local vortex frame. Compared to Eq.~(\ref{eq:LinearMagnetizationFunction}) the number of expansion coefficients in Eq.~(\ref{eq:LinearVortexModel}) has been reduced to two. A positive $m_1$ indicates a counterclockwise (CCW) or right-handed sense of rotation, while a negative $m_1$ corresponds to a clockwise (CW) or left-handed sense of rotation. We note that for a micromagnetic Hamiltonian that contains the isotropic exchange interaction, magnetic anisotropy, the Zeeman, and magnetodipolar interaction, there exists no preference for CCW or CW vortex rotation senses in the particles. CCW and CW vortices appear with equal probability so that the chiral function averages to zero (see below). However, by including the Dzyaloshinskii-Moriya interaction (DMI), which breaks space-inversion symmetry, chirality selection takes place and leads to a nonzero chiral function~\cite{evelynprb2024}.

Equation~(\ref{eq:LinearVortexModel}) models a linear vortex in the local vortex frame. We introduce a $zy$~rotation matrix $\mathbf{R}(\alpha,\beta)$ that transforms the local magnetization $\mathbf{M}'$ into the laboratory frame of reference, where $\alpha$ and $\beta$ denote the (global) polar and azimuthal angles, respectively. The resulting global magnetization vector field is then obtained as:
\begin{align}
\label{eq:LinearVortexModeltransformed}
\mathbf{M}(\mathbf{r}; \alpha, \beta) = \mathbf{R}(\alpha, \beta) \cdot \mathbf{M}'(\mathbf{R}^T(\alpha, \beta) \cdot \mathbf{r}) .
\end{align}
Using Eq.~(\ref{eq:LinearVortexModeltransformed}) in the MNPSE method~\cite{michael2024sm}, we define the ensemble-averaged (dilute) SANS cross sections as:
\begin{align}
\left\langle  \frac{d\Sigma_{\mathrm{sf,\chi}}}{d\Omega}\right\rangle &= \frac{1}{2} \int_{0}^{4\pi} \left[ \frac{d\Sigma_{\mathrm{sf,\chi}}^{\mathrm{CCW}}}{d\Omega} +\frac{d\Sigma_{\mathrm{sf,\chi}}^{\mathrm{CW}}}{d\Omega} \right] \psi(\alpha, \beta) d\Upsilon 
\end{align}
where $d\Upsilon = \sin\alpha d\alpha d\beta$ is the solid-angle differential, and $\frac{d\Sigma_{\mathrm{sf,\chi}}^{\mathrm{CCW}}}{d\Omega}(\mathbf{q}; \alpha,\beta)$ and $\frac{d\Sigma_{\mathrm{sf,\chi}}^{\mathrm{CW}}}{d\Omega}(\mathbf{q}; \alpha,\beta)$ are the SANS cross sections referring to two nanoparticles with the same orientation $(\alpha, \beta)$, but opposite senses of vortex rotation ($m_1^{\mathrm{CCW}} = - m_1^{\mathrm{CW}}$). The function $\psi(\alpha, \beta)$ is a field-dependent probability distribution that models the orientation of both the CCW and CW vortex rotation axes (no distinction between the different polarities); its origin is related to the distribution of the net magnetization vectors of the nanoparticles. For simplicity, we assume a uniform distribution $\psi_{\mathrm{u}}$ on the spherical surface, which is limited by a field-dependent conical opening angle $0^{\circ} \leq \alpha_{\mathrm{c}} \leq 90^{\circ}$. The azimuthally-symmetric distribution is then given by (see~\cite{michael2024sm} for details): 
\begin{align}
\psi_{\mathrm{u}}(\alpha, \beta) = \frac{\Theta(1 - \alpha/\alpha_c)}{2\pi (1 - \cos\alpha_{\mathrm{c}})} ,
\label{eq:UniformDistribution}
\end{align}
where $\Theta(\xi)$ is the Heaviside function. In the fully saturated case ($B_0\rightarrow \infty$) it follows that $\alpha_{\mathrm{c}} \rightarrow 0$, and $\alpha_{\mathrm{c}}$ increases with decreasing applied magnetic field. By inserting Eqs.~\eqref{eq:LinearVortexModel}$-$\eqref{eq:UniformDistribution} into the formalism of the MNPSE method, we obtain the following final expressions for the randomly-averaged 2D spin-flip SANS cross section and chiral function~\cite{michael2024sm}:
\begin{widetext}
\begin{align}
\label{eq:SpinFlip2DSANScrossSection}
\left\langle\frac{d\Sigma_{\mathrm{sf}}}{d\Omega}\right\rangle(q, \theta) &= \frac{W}{8}
[m_0 f(qR)]^2 \times \left[ 12 - (\cos^2\alpha_{\mathrm{c}} +  \cos\alpha_{\mathrm{c}})(3\cos^2(2\theta) + 2\cos(2\theta) + 3) + 4 \cos(2\theta) \right] \\
& + \frac{W}{2} [R m_1 f'(qR)]^2 \times \left[ 3 -  (2 \cos^2\alpha_{\mathrm{c}} + 2 \cos\alpha_{\mathrm{c}} - 1)\cos(2\theta) \right] ,\nonumber \\
\left\langle\frac{d\Sigma_{\chi}^{\mathrm{CCW, CW}}}{d\Omega}\right\rangle(q, \theta) &= \pm W  
 \left[ R m_0 |m_1| f(qR) f'(qR) \cos\theta \right]
   \times\left[4 + \cos^2\alpha_{\mathrm{c}} + \cos\alpha_{\mathrm{c}} - 3 (\cos^2\alpha_{\mathrm{c}} + \cos\alpha_{\mathrm{c}}) \cos^2\theta \right] ,
\label{eq:2DChiralSANScross_sectionCCW}
\end{align}
\end{widetext}
where $W$ is a scaling constant. In Eq.~(\ref{eq:2DChiralSANScross_sectionCCW}) we have separated the chiral function into CCW (``$+$'' sign) and CW (``$-$'' sign) contributions. For the here-considered micromagnetic energy contributions~\cite{michael2024sm} (with no chirality selection taking place), it then follows that
\begin{align}
\left\langle\frac{d\Sigma_{\chi}}{d\Omega}\right\rangle = \frac{1}{2}\left[\left\langle\frac{d\Sigma_{\chi}^{\mathrm{CCW}}}{d\Omega}\right\rangle + \left\langle\frac{d\Sigma_{\chi}^{\mathrm{CW}}}{d\Omega}\right\rangle\right] = 0 .
\end{align}
Figure~\ref{fig2} displays Eqs.~\eqref{eq:SpinFlip2DSANScrossSection} and \eqref{eq:2DChiralSANScross_sectionCCW}. At saturation [Fig.~\ref{fig2}(a,c) and (g,i)], with $\alpha_{\mathrm{c}} = 0^{\circ}$ and $m_1/m_0 = 0$, the spin-flip SANS cross section exhibits the well-known $\sin^2\theta \cos^2\theta$~angular anisotropy. At remanence [Fig.~\ref{fig2}(b,d) and (h,j)], with $\alpha_{\mathrm{c}} = 27^{\circ}$ and $m_1/m_0 = 0.7$, we observe for the spin-flip signal an anisotropy that strongly differs from the saturated case, with maxima for $\theta = 90^{\circ}$. This observation strongly suggests that the magnetization Fourier components are anisotropic, i.e., $\widetilde{M}_{x,y,z} = \widetilde{M}_{x,y,z}(q,\theta)$ (compare to Eq.~(4) in \cite{michael2024sm}). Consequently, to not lose this information, the experimental data analysis should be concentrated on 2D spin-flip data rather than on the azimuthally-averaged 1D data. In micromagnetic simulations of spherical nanoparticles, very similar scattering patterns were observed~\cite{michaeliucrj2023,evelynprb2024}.

Averaging Eq.~(\ref{eq:SpinFlip2DSANScrossSection}) over the angle $\theta$, i.e., $(2\pi)^{-1} \int_0^{2\pi} (...) d\theta$, yields the 1D quantity~\cite{michael2024sm}
\begin{align}
    \langle I_{\mathrm{sf}} \rangle (q) &= \frac{3W}{16} [m_0 f(qR)]^2 (8 - 3 \cos^2\alpha_{\mathrm{c}} - 3 \cos\alpha_{\mathrm{c}})\nonumber \\
&+ 
8 \frac{3W}{16} [m_1 R f'(qR)]^2   .
\label{eq:AngularAveragedSpinFlipAzimuthallyAveragedSANScross sectionVortexSpinStructureModel}
\end{align}
By comparison to Eq.~(\ref{eq:AzimuthallyAveragedSANScrossSectionFirstOrderModel}) we note that the new parameters $m_0$, $m_1$, and $\alpha_{\mathrm{c}}$ in Eqs.~\eqref{eq:SpinFlip2DSANScrossSection} and \eqref{eq:2DChiralSANScross_sectionCCW} are related to the coefficient ratio $I_{\mathrm{sf}}^{1}/I_{\mathrm{sf}}^{0}$ as follows:
\begin{align}
\iota &= \frac{I_{\mathrm{sf}}^{1}}{I_{\mathrm{sf}}^{0}} = \frac{8 m_1^2 R^2}{m_0^2 (8 - 3\cos^2\alpha_{\mathrm{c}} - 3\cos\alpha_{\mathrm{c}})} ,
\end{align}
which emphasizes the importance of the vortex-axes distribution function. The angle $\alpha_{\mathrm{c}}$ may be obtained from the analysis of (preferentially 2D) experimental spin-flip SANS data (compare to Fig.~\ref{fig2}(d) and (j) and the video clip in~\cite{michael2024sm}).

\textit{Conclusion.} In this Letter we have demonstrated that the linear MNPSE approach captures the main effects in the spin-flip SANS cross section and pair-distance distribution function stemming from dilute assemblies of spherical nanoparticles exhibiting vortex-type spin textures. A crucial insight is that the linear functionality represents the most important contribution to the magnetic neutron scattering cross section. Based on the specific case of a linear vortex model, we have derived analytical expressions for the 2D and 1D spin-flip and chiral cross sections of an ensemble of randomly oriented vortex-carrying nanoparticles. The maximum of the spin-flip scattering intensity and the zero of the pair-distance distribution function appear, respectively, at momentum transfer $q_{\mathrm{max}} \cong 2.50/R$ and position $r_{\mathrm{z}} \cong 1.07 R$, where $R$ denotes the radius of the spherical nanoparticles. The analytical predictions, which enable e.g.\ the determination of the field-dependent conical opening angle $\alpha_{\mathrm{c}}$ of the vortex-axes distribution from experimental data, are in very good agreement with the results of micromagnetic simulations. The chiral SANS cross section is sensitive to the vortex rotation sense, but in a many-particle system with no chirality selection, it averages to zero (as expected). A candidate for chirality selection is the DMI interaction that breaks space-inversion symmetry.

We acknowledge financial support from the National Research Fund of Luxembourg (AFR Grant No.~15639149, CORE Grant DeQuSky, and PRIDE MASSENA Grant).

%\bibliography{BIB}

%apsrev4-2.bst 2019-01-14 (MD) hand-edited version of apsrev4-1.bst
%Control: key (0)
%Control: author (72) initials jnrlst
%Control: editor formatted (1) identically to author
%Control: production of article title (-1) disabled
%Control: page (0) single
%Control: year (1) truncated
%Control: production of eprint (0) enabled
%

\end{document}